\documentclass[conference,compsoc]{IEEEtran}
\ifCLASSOPTIONcompsoc
  \usepackage[nocompress]{cite}
\else
  \usepackage{cite}
\fi
\ifCLASSINFOpdf
\else
\fi

\usepackage{booktabs}
\usepackage{amsmath}
\usepackage{amsmath,amssymb,amsfonts}
\usepackage{algorithm}
\usepackage{algpseudocode}
\usepackage{graphicx}
\usepackage[hidelinks]{hyperref}

\usepackage{amsthm}
\newtheorem{theorem}{Theorem}
\usepackage{comment}

\usepackage{xcolor}
\usepackage{listings}
\lstdefinestyle{tcstyle}{
  backgroundcolor=\color{gray!10},
  basicstyle=\ttfamily\small,
  keywordstyle=\color{blue}\bfseries,
  commentstyle=\color{green!60!black}\itshape,
  stringstyle=\color{orange},
  identifierstyle=\color{black},
  breaklines=true,
  breakatwhitespace=true,
  frame=single,
  framerule=0.5pt,
  rulecolor=\color{gray!50},
  numbers=left,
  numberstyle=\tiny\color{gray},
  numbersep=6pt,
  xleftmargin=5pt,
  xrightmargin=5pt,
  showstringspaces=false,
  columns=fullflexible
}

\def\TC#1{\textsc{Trust}\-\textsc{Checkpoints}}
\newcounter{remark}[section]

\def\editingmarks{}
\ifx\editingmarks\undefined
\newcommand{\myremark}[3]{}
\else
\newcommand{\myremark}[3]{
\refstepcounter{remark}
\[
\left\{
\sf
\parbox{0.8\columnwidth}
{
{\bf {#1}'s remark~\arabic{section}.\arabic{remark}:}
{#3}
}
\right.
\]
\marginpar{\bf {#2}.~\arabic{section}.\arabic{remark}}
}
\fi
\def\point#1{{\bf #1:}}

\def\colfig#1{\includegraphics[width=0.8\linewidth]{#1}}

\begin{document}

\title{\TC{}: Time Betrays Malware for Unconditional Software Root of Trust}

\author{
\IEEEauthorblockN{Friedrich Doku}
\IEEEauthorblockA{McCormick School of Engineering\\
Northwestern University\\
Evanston, Illinois 60208\\
Email: friedy@u.northwestern.edu}
\and
\IEEEauthorblockN{Peter Dinda}
\IEEEauthorblockA{McCormick School of Engineering\\
Northwestern University\\
Evanston, Illinois 60208\\
Email: pdinda@northwestern.edu}}

\maketitle

\begin{abstract}
Modern IoT and embedded platforms must start execution from a known trusted state to thwart malware, ensure secure firmware updates, and protect critical infrastructure. Current approaches to establish a root of trust depend on secret keys and/or specialized secure hardware, which drives up costs, may involve third parties, adds operational complexity, and relies on assumptions about an attacker's computational power. In contrast, \TC{} is the first system to establish an unconditional software root of trust based on a formal model—without relying on secrets or trusted hardware. Developers capture a full-system checkpoint and later roll back to it and prove this to an external verifier. The verifier issues timing-constrained, randomized $k$-independent polynomial challenges (via Horner's rule) that repeatedly scan the fast on-chip memory in randomized passes. When malicious code attempts to persist, it must swap into slower, unchecked off-chip storage, causing a detectable timing delay.

Our prototype for a commodity ARM Cortex-A53-based platform validates 192~KB of SRAM in $\sim$10~s using 500 passes, sufficient to detect single-instruction persistent malware. The prototype then seamlessly extends trust to DRAM. Two modes---fast SRAM-bootstrap and comprehensive full-memory scan---allow trade-offs between speed and coverage, demonstrating reliable malware detection on unmodified hardware.
\end{abstract}

\section{Introduction}
\label{sec:intro}

As embedded systems and Internet of Things (IoT) devices proliferate across critical infrastructure, industrial automation, medical instrumentation, and consumer electronics, ensuring their integrity has never been more crucial. These devices often operate unattended, control physical processes, and handle sensitive data, making them compelling targets whose compromise can lead to severe safety, privacy, or financial consequences.

A root of trust ensures that devices begin execution from a known-good state, free from persistent malware. It is the foundation upon which all subsequent security guarantees rest: if the root is compromised, no downstream software can be trusted. Establishing this root is critical for secure boot, firmware validation, and continuous integrity monitoring in distributed deployments~\cite{madea, sbap}.  

Existing approaches to establish a root of trust generally depend on (1) public-key cryptography and secure key storage, (2) specialized hardware modules such as TPMs or HSMs, or (3) complexity assumptions (e.g., RSA or discrete logarithm hardness)~\cite{intelsgx, litehax, tpm2, nunes2022VRASED}. All three dependencies pose challenges in low-cost, resource-constrained environments and may be undermined by future quantum adversaries~\cite{shor}. Furthermore, existing software-based verification schemes can check code integrity but cannot detect malware hidden in system state~\cite{Gligor2021UROT}.

More fundamentally, these cryptographic roots of trust suffer from an irreversibility problem: once secrets are compromised, there exists no cryptographically sound method to determine whether a device remains infected. This creates a fundamental asymmetry where defenders must protect secrets indefinitely, while attackers need succeed only once. Current approaches cannot answer the critical post-compromise question: \emph{``Is this device still compromised?''}

This inability to verify cleanliness after potential compromise has severe practical consequences. Organizations facing sophisticated adversaries must treat any device with potentially leaked credentials as permanently untrusted, leading to costly hardware replacement cycles. Worse, advanced persistent threats can maintain presence even through credential rotation, as verification schemes cannot detect malware residing in system state rather than code.

We observe that while cryptographic protocols excel at preventing unauthorized access, they fundamentally cannot detect unauthorized presence once access is obtained. This limitation stems from Shannon's perfect secrecy: an adversary with complete knowledge of all secrets becomes cryptographically indistinguishable from legitimate users. Breaking this symmetry requires introducing an asymmetry the adversary cannot replicate---not in computational power, but in physical constraints.

\TC{} exploits a simple physical reality: computation takes time, and this time cannot be hidden. By forcing devices to perform carefully crafted computations from known checkpoints while precisely measuring execution timing, we can detect any active malware through unavoidable timing perturbations. Unlike cryptographic approaches that fail catastrophically upon key compromise, \TC{}'s timing-based verification degrades gracefully---an adversary who steals keys gains no advantage in hiding their computational footprint.

Specifically, \TC{} establishes a \emph{physically-grounded} root of trust by restricting device state to verified checkpoints and measuring the precise timing of randomized polynomial computations. Any deviation from expected timing behavior---whether from resident malware, rootkits, or hardware implants---becomes detectable with probability approaching certainty as measurements increase. This provides, for the first time, a mechanism to verify device cleanliness that remains effective even against adversaries with complete cryptographic knowledge.

\TC{} builds on the theoretical foundation established by Gligor et al.~\cite{Gligor2021UROT,gligor-whats-necessary-for-practical-system, gligor-software-only-root-of-trust-fact-and-fiction}, taking their theoretical approach and applying it on real commodity hardware. Our current prototype represents the first working implementation of unconditional software root of trust, bridging the gap between theory and practice. Our experience building \TC{} reveals both the challenges and opportunities in making timing-based verification practical, providing concrete insights for future systems that could achieve better performance through lower-jitter channels or hardware-assisted timing isolation.

The heart of \TC{} is a $k$-independent randomized-polynomial evaluation via Horner's rule over the entire checkpointed memory. By repeatedly scanning the fast on-chip memory in multiple randomized passes, any attacker attempting to hide malicious code must swap it to slower off-chip storage (e.g., MMC or SPI flash). The accumulated swap time incurs a measurable latency penalty detectable by our microsecond-resolution timing mechanism. We can thus detect when malware persists, preventing establishment of trusted state. Conversely, successful checkpoint restoration within the expected time bound guarantees that the system has returned to a malware-free state, establishing our software root of trust.

Our contributions are as follows:
\begin{itemize}
\item We present the \TC{} methodology (\S\ref{sec:methodology}), the first truly unconditional mechanism for establishing software root of trust that leverages $k$-independent randomized polynomials to confine execution to a small set of trusted memory snapshots and verify malware-free rollback, without relying on stored secrets or specialized hardware. 
\item We describe the design and implementation of the \TC{} prototype for commodity ARM hardware (\S\ref{sec:prototype}). It provides a simple user-level API for capturing checkpoints and verifying malware-free rollback to establish software root of trust.
\item We provide detailed implementation guidance to help future system designers avoid pitfalls when building unconditional verification systems on their target platforms (\S\ref{sec:practical-challenges}).
  \item We evaluate our prototype using the minimum possible persistent malware, a single instruction. Our prototype can detect this attack with a near zero false negative rate, while only incurring a tiny false positive rate when no attack exists.

\end{itemize}
Our results demonstrate that \TC{}, without relying on cryptographic keys or trusted hardware, can reliably detect an injected payload that consists of just a single instruction.  Our design and software will be made available on publication of this paper.

\section{Background and Related Work}
\label{sec:bg}
Establishing a root of trust has become fundamental to system security, ensuring devices begin execution from a known-good state free from persistent malware. This is especially vital in the Internet of Things (IoT), where fleets of resource-constrained embedded devices face adversaries with widely varying capabilities. The root of trust forms the foundation upon which all security guarantees are built—if compromised, no downstream software can be trusted.

Traditional approaches to establishing root of trust rely on three main strategies: (1) immutable hardware roots such as mask ROM or one-time programmable memory that cannot be modified by attackers, (2) cryptographic mechanisms using secret keys stored in protected hardware enclaves, or (3) specialized trusted hardware modules like TPMs, HSMs, or TEEs that provide isolated execution environments. While these approaches have seen substantial adoption and standardization (e.g., in Intel SGX~\cite{intelsgx}), they pose challenges for devices with extremely limited computational and memory resources. 

Beyond establishing the initial root of trust, a critical challenge remains: verifying that a device has successfully returned to its trusted state after potential compromise. Attestation protocols have evolved to address this verification challenge, enabling a trusted verifier to assess whether a device has rolled back to a clean state. However, existing attestation schemes inherit the same dependencies as their underlying root of trust mechanisms, requiring either cryptographic secrets, trusted hardware, or computational hardness assumptions that may be undermined by future quantum adversaries.

\subsection{Hardware-Based Attestation}

Early attestation protocols were primarily hardware-based, building on Trusted Platform Modules (TPMs) and secure enclaves like Intel SGX or ARM TrustZone~\cite{intelsgx, arm-trustzone, tpm2}. These solutions can provide strong roots of trust but require developers to spend substantial time porting their applications and device drivers to them~\cite{driverlet, nickgordon, ldr}. Furthermore, hardware-centric methods often introduce additional cost, complexity, and reliance on external parties to provision secrets at manufacturing time. Finally, hardware vulnerabilities~\cite{tpmfail} are both extremely challenging to address after the fact and an obvious target for attackers.

\subsection{Software-Only Attestation}

Software-only attestation schemes such as Pioneer and SWATT~\cite{Pioneer, SWATT} avoided dedicated security hardware by embedding timing checks and self-checksum loops directly into the prover's code. These methods embed a self-checking loop that reads memory in a pseudo-random order and measures execution time to detect modifications. While practical on legacy embedded platforms, these approaches remain largely heuristic: they offer no formal lower bound on how much adversarial work can be hidden, and they can be defeated by local adversaries who replay stale responses, reorder computations, or exploit predictable memory regions.

Armknecht et al. formalized this space in a generic framework, identifying precise conditions (e.g., memory incompressibility, oracle-modeled primitives, time bounds) for provable security~\cite{Armknecht}, but their analysis stops short of a concrete, keyless implementation on real hardware.

\subsection{Hybrid and Hypervisor-Based Approaches}

Hybrid attestation protocols, which combine minimal hardware trust anchors with lightweight software checks, represent a middle ground. Projects like SMART and VRASED~\cite{Defrawy2012SMARTSA, nunes2022VRASED} have shown that by using a small hardware root of trust and formally verified code, it's possible to provide strong security guarantees even on resource-limited devices. This is further exemplified in protocols like SeED~\cite{SeED} or ERASMUS~\cite{ERASMUS}, which use periodic self-measurement and loosely synchronized clocks to achieve efficient, scalable attestation without expensive hardware dependencies.

In cloud and enterprise environments, hypervisor-assisted schemes have emerged. XSWAT~\cite{xswat} adapts timing-based attestation to cloud hypervisors by integrating Xen kernel modules, Intel's Last Branch Record, and SHA-1 checksums, eliminating the need for TPMs while defending against time-of-check to time-of-use (TOCTOU) and multi-core races. Checkmate~\cite{kovah} takes a Windows-centric approach, measuring end-to-end network RTTs via an NDIS intermediate driver to detect code-integrity violations over enterprise networks with minimal overhead. These systems achieve impressive performance in their domains but still depend on complex software stacks, public-key infrastructure, or centralized baselining.

\subsection{Memory-Based Detection Schemes}

Jakobsson and Johansson's ``memory-printing'' approach~\cite{jako} detects active malware by exploiting the timing gap between RAM and slower storage like flash, relying on the assumption that malware incurs detectable delays when accessing off-chip memory. However, it offers only heuristic guarantees and lacks formal bounds on adversarial effort or detection confidence.

\subsection{Unconditional Security}

Unconditional security requires no on‐device secrets, trusted hardware modules, or special instructions (e.g., TPMs, ROMs, SGX), and does not assume any bound on the adversary’s computational power \cite{Gligor2021UROT}. The external verifier stores no secrets, executes no code on the device being challenged, and confers no additional capabilities. 

This stands in contrast to \emph{conditional} security approaches that rely on unproven computational assumptions (e.g., factoring or discrete logarithm hardness), trusted third parties, or pre-shared cryptographic material.
Unconditional security solutions offer several fundamental advantages over conditional ones:
\begin{itemize}
\item \textbf{Independence from third parties:} They require no security mechanisms, protocols, or external parties whose trustworthiness is uncertain, such as secret keys installed in hardware by manufacturers.
\item \textbf{Provable adversary limitations:} They limit any adversary's chance of success to provably low probabilities determined by the defender, giving defenders undeniable mathematical advantage.
\item \textbf{Computational independence:} They remain secure regardless of the adversary's computing power or technology, including quantum computers.
\end{itemize}
Unconditional security systems derive their guarantees from \emph{physical properties} rather than \emph{computational assumptions}; the verifier needs only the physical device specifications, such as memory speeds and CPU characteristics. For embedded and IoT devices, this eliminates dependence on heavyweight cryptographic libraries, complex key management infrastructure, or trust in hardware manufacturers' key provisioning processes.

At its core, unconditional security transforms the trust model from computationally ``hard-but-not-impossible'' to ``provably
impossible''. Security depends solely on measurable physical phenomena, such as memory access timing, hardware randomness sources, or communication channel properties rather than unproven mathematical conjectures about computational difficulty.

 Gligor and Woo's seminal work formalizes this approach by defining a concrete Word Random Access Machine (cWRAM) model and introducing $k$-independent randomized-polynomial primitives with rigorous space–time optimality guarantees~\cite{Gligor2021UROT}. Their theoretical framework proves that any adversarial deviation from optimal polynomial evaluation must incur detectable additional work, but stops short of demonstrating a concrete system.
 
\TC{} bridges this theory-practice gap by realizing unconditional security on commodity hardware. Using only a source of physical randomness (for polynomial coefficients) and the externally measurable time required to complete the challenge, we demonstrate that unconditional security is achievable, providing information-theoretic guarantees without cryptographic assumptions, pre-shared secrets, or trusted hardware vendors.

\subsection{Randomized Polynomials: Foundation and Security Properties}
\label{randomized_poly}

\begin{figure}
\begin{centering}
    \begin{tabular}{ll}
       \hline
       Symbol & Meaning \\
       \hline
       $v = (v_0, v_1, \ldots, v_d)$ & Memory content being challenged \\
       $s_i$ & Polynomial coefficients \\
       $r_0, r_1, \ldots, r_{k-1}$ & Random values \\
       $x \in \mathbb{Z}_p$ & Random evaluation point \\
       $\pi$ &  Permutation \\
       $\oplus$ & Bitwise XOR operation \\
       \hline
     \end{tabular} \\
   \end{centering}
     \caption{Symbols used in this paper and their meanings.}
     \label{fig:symbols}
\end{figure}

In the remainder of Section~\ref{randomized_poly} we quote Gligor~\cite{Gligor2021UROT}.
\TC{} employs \emph{randomized polynomials} as its core cryptographic primitive for establishing a software root of trust. These polynomials provide provable security guarantees against adversarial manipulation. Our paper uses a range of symbols starting from this point. Figure~\ref{fig:symbols} presents a guide.

\subsubsection{Mathematical Definition}

A randomized polynomial $H_{d,k}(\cdot)$ of degree $d$ over the finite field $\mathbb{Z}_p$ is defined as:
\label{coff}
\begin{equation}
H_{d,k}(v) = \sum_{i=0}^{d} (v_i \oplus s_i) \times x^i \pmod{p}
\end{equation}
where:
\begin{itemize}
    \item $v = (v_0, v_1, \ldots, v_d)$ represents the memory content being challenged
    \item $s_i = \sum_{j=0}^{k-1} r_j \times (i+1)^j \pmod{p}$ are the polynomial coefficients
    \item $r_0, r_1, \ldots, r_{k-1}$ are $k$ random values chosen uniformly from $\mathbb{Z}_p$
    \item $x \in \mathbb{Z}_p$ is a random evaluation point
    \item $\oplus$ denotes the bitwise XOR operation
\end{itemize}

\subsubsection{Key Security Properties}

\paragraph{$k$-wise Independence}
The coefficients $s_i$ exhibit $k$-wise independence, meaning any subset of $k$ coefficients appears uniformly random and independent. This property ensures that an adversary cannot predict coefficient values even with partial knowledge of up to $k-1$ coefficients. Formally:

\begin{theorem}[k-wise Independence]
For any distinct indices $i_1, i_2, \ldots, i_k$ and any values $a_1, a_2, \ldots, a_k \in \mathbb{Z}_p$:
\[
\Pr[s_{i_1} = a_1 \wedge s_{i_2} = a_2 \wedge \cdots \wedge s_{i_k} = a_k] = \frac{1}{p^k}
\]
\end{theorem}

This independence prevents an adversary from using knowledge of some memory locations to predict the polynomial's behavior at other locations.

\paragraph{Second Pre-image Resistance}
The randomized polynomial construction provides strong collision resistance properties:

\begin{theorem}[Collision Resistance]
For any $x \in \mathbb{Z}_p$ and $y \neq x$:
\[
\Pr[H_{d,k}(y) = H_{d,k}(x)] \leq \frac{1}{p-1}
\]
\end{theorem}

This bound ensures that finding two different memory configurations that produce the same result is computationally infeasible for large $p$.

\paragraph{Space-Time Optimality}
The randomized polynomial $H_{d,k}(\cdot)$ achieves provable space-time optimality in the cWRAM model, which captures realistic instruction-level execution and memory constraints. It closely reflects real hardware by incorporating a fixed word size, general-purpose instruction sets with multiple addressing modes, and support for I/O operations, caches, virtual memory, and multiprocessors. Unlike idealized models, cWRAM captures the concrete instruction-level and memory-access behavior of real systems~\cite{Gligor2021UROT}.

\begin{theorem}[Concrete Bounds in cWRAM]
Any adversarial evaluation of $H_{d,k}(\cdot)$ that returns a correct result must use:
\begin{itemize}
    \item At least $k + 22$ words of memory, and
    \item At least $(6k - 4) \times 6d$ clock cycles,
\end{itemize}
except with probability at most $\frac{3}{p}$.
\end{theorem}

Horner's rule is used to establish concrete lower bounds on the work required to evaluate a polynomial. In infinite fields (like the real numbers), it has been proven that Horner's rule is uniquely optimal, meaning no other method can use fewer basic operations (addition, subtraction, multiplication, or division) \cite{horner}. Any correct algorithm for evaluating a general polynomial must perform at least as much arithmetic as Horner’s rule does.

While this optimality does not hold in general for finite fields \cite{elia_polynomial_2012}, where alternative strategies may reduce cost by exploiting algebraic structure, the cWRAM model restores Horner’s unique optimality by simultaneously minimizing both time and space. In this model, every instruction and memory access is explicitly accounted for, and any deviation from Horner’s structure incurs a measurable cost in either execution time or memory footprint. This makes Horner’s rule not just efficient, but provably minimal within the cWRAM framework.

These bounds were derived analytically under the cWRAM model by modeling every instruction and memory access involved in the Horner-rule evaluation of $H_{d,k}(\cdot)$. While these bounds are tight and formally proven in the cWRAM setting, our goal is to assess how well these guarantees hold on real hardware, where unmodeled effects, such as microarchitectural behavior, timing jitter, and physical variability can challenge assumptions made in abstract models. Understanding the correspondence between theoretical bounds and real-world execution is critical for establishing unconditional software root of trust through timing-based verification.

\section{Methodology}
\label{sec:methodology}

\begin{figure}
  \centerline{\colfig{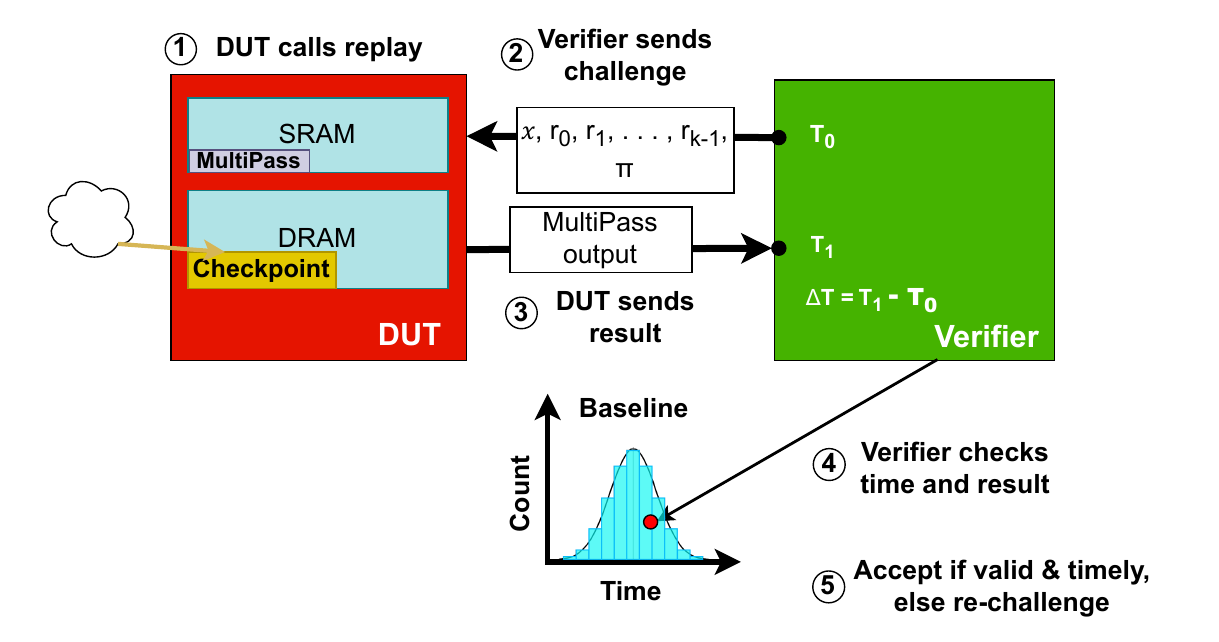}}
  \caption{\TC{} general architecture.} 
\label{fig:archx}
\end{figure}

\begin{figure}
  \centering
  \colfig{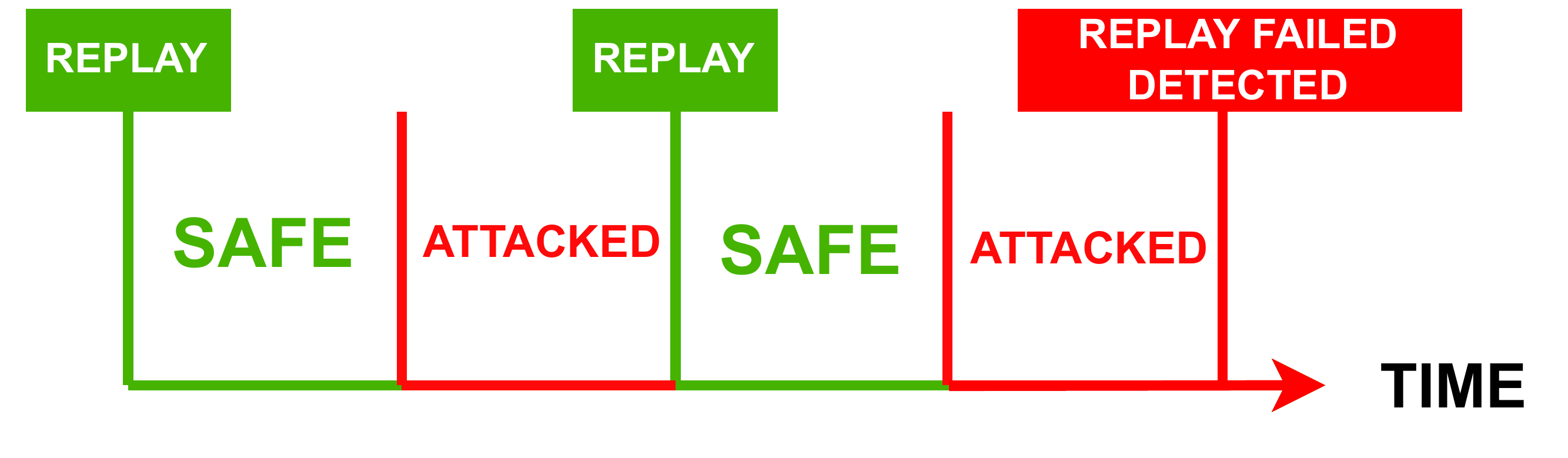}
  \caption{This timeline illustrates how a device can recover from malware by replaying previously recorded safe checkpoints. At each ``REPLAY'' point, the system restores a snapshot of CPU registers, on-chip SRAM, and selected DRAM regions, returning to a previously verified safe state. }
  \label{fig:adversary-model}
\end{figure}

We begin by presenting the overall methodology and system architecture, followed by a detailed breakdown of each component of \TC{}. Figure~\ref{fig:archx} offers a visual overview, while Figure~\ref{fig:adversary-model} illustrates how checkpoints are used to recover the system during execution.

\subsection{Overview}
\TC{} consists of three principal components: the \emph{developer interface}, the \emph{external verifier}, and the \emph{device under test} (DUT).  During normal operation, developers insert lightweight API calls into their programs to record and restore trusted system states:
\begin{itemize}
  \item \texttt{checkpoint\_record()} captures a snapshot of CPU registers, on-chip SRAM, and selected regions of DRAM, and stores it in a reserved memory region
  \item \texttt{checkpoint\_replay()} restores the device into a previously recorded checkpoint, overwriting any intervening modifications.
\end{itemize}

At challenge time, a \emph{challenge processor} generates a fresh randomized‐polynomial challenge and sends it over a secure channel to the external verifier. The external verifier is a microcontroller connected over a low‐variability link. It relays the challenge to the DUT, signals the DUT to restore the checkpoint, and then begins high‐precision timestamping.  The DUT executes the polynomial evaluation by scanning the selected memory hierarchy (SRAM only or both SRAM and DRAM) including CPU registers in multiple randomized passes, including it's challenge program. On completion, the DUT emits the result value, which the microcontroller immediately timestamps and forwards back to the challenge processor.

The challenge processor validates the final polynomial result by comparing its execution time against a known baseline. Malware that swaps to slower memory to evade detection introduces measurable latency deviations.

By building on keyless, information‐theoretic primitives and an out-of-band timing channel, this architecture delivers end-to-end guarantees without relying on stored secrets, trusted hardware modules, or complexity assumptions.  The two provided modes—\emph{SRAM‐only} and \emph{full‐memory}—allow deployers to balance challenge scope against runtime overhead, making \TC{} suitable for a wide range of embedded platforms.

\subsection{Threat Model and Trust Assumptions} 
In our threat model, the verifier seeks to establish that a device it physically possesses, such as an embedded system, IoT node, or controller, is free from malware without trusting any software running on it. \TC{} achieves this without on-device secrets or specialized on-chip secure hardware (e.g., TPM/TEE/HSM), but does rely on specific trust assumptions detailed below.

\subsubsection*{Trust Assumptions}
\begin{itemize}
  \item \textbf{External Verifier:} A trusted external microcontroller connected via a bounded-jitter link performs all timing measurements and challenge generation. This verifier and its physical connection are assumed tamper-proof.
\item \textbf{Timing Stability:} The device must guarantee bounded timing noise, achieved through fixed clock frequencies, disabled DVFS, and quiesced peripherals during the challenge.

  \item \textbf{DMA Containment:} \TC{} remains secure only on systems where all DMA-capable peripherals are either fully trusted or can be verifiably disabled. We outline three defense strategies to enforce this requirement (see Section~\ref{sec:dma}).
  \item \textbf{Baseline Profiling:} During a trusted setup phase, the verifier empirically profiles the device's timing behavior to establish a baseline for detecting deviations.
  \item \textbf{Public Checkpoints:} The memory snapshots (checkpoints) themselves are public information requiring no confidentiality.
\end{itemize}

Under these assumptions, our security is unconditional with respect to attacker computational power.

\subsubsection*{Adversary Capabilities}
\begin{itemize}
  \item \textbf{Persistent Malware:} Implant malware that survives power cycles, secure/trusted boot, and firmware reflashing.
  \item \textbf{System Control:} Modify the system state at any software layer—firmware, kernel, or application—but \emph{not} hardware.
  \item \textbf{Adaptive Code Modification:} Alter challenge-processing code on-the-fly, e.g., to shortcut or replay portions of the polynomial evaluation.
  \item \textbf{I/O Channel Access:} Read from and write to the DUT's link, attempting to spoof or delay timestamped messages.
  \item \textbf{Baseline Awareness:} The adversary possesses the device specifications and knows the expected execution time of the challenge under normal conditions.
\end{itemize}

\subsubsection*{Adversary Limitations}
\begin{itemize}
  \item \textbf{External Components:} Cannot tamper with the external verifier microcontroller or compromise the bounded-jitter serial link.
  \item \textbf{Immutable Hardware:} Cannot modify the device's physical hardware—only its firmware/software.
  \item \textbf{Peripheral Control:} Cannot prevent hardware-enforced peripheral resets or DMA quiescence when properly configured.
  \item \textbf{Randomness Protection:} Cannot predict random nonces issued by the external verifier.
  \item \textbf{Denial of Service:} We do not defend against pure DoS attacks (power-cycling, link-jamming, etc.).
\end{itemize}

\subsection{Checkpointing}
\label{sec:checkpointing}

Before any field device can be challenged, the owner must first generate and distribute a trusted checkpoint.  This provisioning consists of two main steps: \emph{baseline calibration \& checkpoint capture} on a trusted reference device, and \emph{replay} on each target device.

\point{A. Baseline Calibration \& Checkpoint Capture}  
On a secure reference device (e.g.\ in the factory or lab), the owner will:
\begin{enumerate}
  \item \textbf{Configure Known‐Good State.}  
    Boot the device into the desired firmware/OS configuration (hypervisor, kernel, applications), disable non‐essential services, and verify correct operation.
  \item \textbf{Calibrate Timing.}  
    Execute the \textsc{MultiPass} randomized‐polynomial routine (see Algorithm~§\ref{alg:randMultiPass}) repeatedly (e.g.\ 50–100 trials).  The external verifier MCU records start/stop timestamps to capture the empirical distribution and serial correlation.  

  \item \textbf{Record Trusted Checkpoint.}  
    Invoke
    \begin{verbatim}
      checkpoint_record();
    \end{verbatim}
    which snapshots:
    \begin{itemize}
      \item CPU general‐purpose registers and control state (e.g.\ hypervisor context),
      \item the entire on‐chip SRAM image,
      \item only the selected DRAM memory of interest,
      \item any other critical data (device‐tree blobs, kernel image, application binaries).
    \end{itemize}
    The snapshot is stored in a reserved part of DRAM. The checkpoint data can be read from anywhere as long as it is loaded into DRAM. 
  \item \textbf{Export \& Distribute.}  
    Package the checkpoint image together with its baseline data, and distribute it to each DUT with the same device specifications. The checkpoint image can be made public. The connectivity between the DUT and the verifier can be any channel as long as the channel has low variability (jitter). If the link's timing fluctuations are too large, it becomes difficult to distinguish between natural transmission delays and adversarial behavior. 
\end{enumerate}

\point{B. Stateless Restoration on Target DUTs}  
Each Device-under-test (DUT) establishes its software root of trust by loading the owner-provided checkpoint package into a reserved part of DRAM and invokes \texttt{checkpoint\_replay()}, which restores CPU registers, on-chip SRAM, and the designated DRAM regions to their checkpointed values. It also resets all other volatile state (including caches and peripherals) back to the trusted checkpoint. Once \texttt{checkpoint\_replay()} returns, the DUT retains no memory of any execution preceding the checkpoint and is immediately ready to prove successful rollback through the timing challenge.

\subsection{Multi-Pass Evaluation Protocol}
\label{multipass_algo}

\TC{} uses multiple passes of the Horner evaluation of $k$-independent randomized polynomials to strengthen detection, forcing adversaries to swap out their added code on each pass, incurring cumulative delays that are externally observable.

\begin{algorithm}[ht]
\caption{\textsc{MultiPass} Randomized Polynomial Evaluation}\label{alg:randMultiPass}
\begin{algorithmic}[1]
\Require 
  Memory snapshot $v[0\ldots d-1]$, \\
  Random values $r_0, r_1, \ldots, r_{k-1}$, \\
  field element $x \in \mathbb{Z}_p$, \\
  prime modulus $p$, \\
  permutation seed $\mathit{seed}$, \\
  number of passes $P$ \\
\Ensure 
  Final accumulator $\mathit{result}$
\State $d \gets \mathrm{length}(v)$
\State $\pi \gets \mathrm{pseudorandom\_permutation}(d, \mathit{seed})$
\State $\mathit{result} \gets 0$
\For{$\mathit{pass} \gets 0$ to $P-1$}
  \For{$i \gets 0$ to $d-1$}
    \State $\mathit{idx} \gets \pi[d - 1 - i]$
    \State $\mathit{coeff\_idx} \gets \mathit{pass} \cdot d + \mathit{idx}$
    \State $s_i \gets \sum_{j=0}^{k-1} r_j \times (\mathit{coeff\_idx}+1)^j \bmod p$ 
    \State $\ell \gets v[\mathit{idx}] \oplus s_i$
    \State $\mathit{result} \gets (\mathit{result} \times x + \ell) \bmod p$
  \EndFor
\EndFor
\State \Return $\mathit{result}$
\end{algorithmic}
\end{algorithm}

\textsc{MultiPass} evaluates a $k$-independent randomized polynomial over a memory snapshot using Horner's method, repeated across multiple passes to amplify detection robustness. The input $v[i]$ refers to the $i$th memory word and is accessed in a fixed pseudorandom order determined by a permutation $\pi$. 

Critically, the coefficients are computed on-demand rather than stored in memory. For each pass $p$ and index $i$, a unique coefficient $s_i$ is computed on the fly using the $k$ random values $r_0, r_1, \ldots, r_{k-1}$ provided by the verifier. Following the formula from Section~\ref{coff}, each coefficient is derived as $s_i = \sum_{j=0}^{k-1} r_j \times (i+1)^j \bmod p$, immediately used in the XOR operation, and then discarded. At no point is an array of coefficients materialized in memory. The only persistent state consists of the $k$ random values and loop variables, all of which fit entirely in registers. This design is crucial for security: the registers are fully occupied during execution, and any attempt by an attacker to commandeer them would require additional instructions or memory traffic, introducing detectable delays.

The accumulator is updated via Horner's rule using a shared field element $x \in \mathbb{Z}_p$, with all computations modulo a prime $p$. Each pass uses a distinct coefficient index range (determined by $\mathit{pass} \cdot d + \mathit{idx}$), ensuring that coefficients are never reused across passes and preventing adversarial alignment of memory layouts.

If an attacker wants to hide malware, they must evict some legitimate memory content to make room. Since the challenge runs multiple passes over memory in random order, the attacker faces an impossible dilemma:

\begin{itemize}
\item To persist across passes, their malware must stay in the checked memory
\item But to compute the correct result, they need the original code they evicted
\item So they must constantly swap: original code out, malware in; then malware out, original code back in, etc.
\end{itemize}

Each swap requires accessing unchecked, slower off-chip memory (DRAM, flash, etc.), adding measurable delays. Over hundreds of passes, these tiny delays accumulate into a timing difference our external verifier can detect. The attacker cannot avoid this, they need both their malware and the original code, but only have space for one at a time. Our space bound doesn't have room for malware.

\subsubsection{Critical Design Features}
\label{sec:randomized-access}

A critical challenge in memory verification arises when adversaries exploit predictable or highly compressible regions, e.g., zero-filled pages or repeated patterns, to shortcut genuine memory fetches by replaying precomputed values. An adversary who knows that certain regions are trivial or compressible might skip the actual memory reads during the challenge and return the correct final result in less time than the honest DUT. To defeat such compression-based evasion strategies, \textsc{MultiPass} accesses memory in a pseudorandom order, ensuring that every word must be fetched from its actual physical location.

This gives us:

\begin{itemize}
  \item \emph{Unpredictability:}  Without the seed, guessing the next address has probability at most \(1/(d-i)\) at step \(i\).
  \item \emph{Uniform coverage:}  Every word is accessed exactly once, so no region can be omitted.
\end{itemize}

We do not perform on-the-fly histogramming or statistical correction in the protocol itself. Instead, if deployers detect low-entropy regions in their checkpointed memory (e.g., via simple histogram or compression-ratio measurements), they can trivially inject randomness by filling unused memory with random values. 

Our randomized permutation forces adversaries to access memory in an unpredictable order. While an adversary might compress sequential zero-filled pages offline, they cannot predict which memory word will be requested next during the challenge. Each access requires fetching the actual data from its physical location. Compression only helps if you know the access pattern in advance. Our permutation depends on the verifier's random seed, which is unknown to the adversary in advance.

\medskip
By combining \textsc{MultiPass} (Algorithm~\ref{alg:randMultiPass}) with a one‐time entropy check, \TC{} delivers a lightweight, deployer‐driven defense against all compression‐based optimizations, while still preserving our core guarantee.

\subsection{Space–Time Optimality and Forced Storage Swaps}
\label{sec:spacetime-optimality}

A key pillar of our design is the space–time optimality of Horner’s‐rule evaluation on a fixed memory footprint.  In the cWRAM model, Gligor and Woo prove that any program which simultaneously minimizes execution time (one multiply–add per coefficient) and working memory (exactly $k+1$ memory words for a degree‐$k$ polynomial) must conform to the unique Horner‐rule layout~\cite{Gligor2021UROT}. Our honest DUT implementation fits exactly within on-chip SRAM and exhibits runtime consistent with the expected performance of Horner-based polynomial evaluation.

\medskip

\noindent\textbf{Adversarial Space Violations.}  
Suppose an attacker attempts to inject malicious code or data alongside the Horner evaluator.  Let $S_{\mathsf{SRAM}}$ denote the total on‐chip SRAM reserved for evaluation and $S_{\mathsf{M}}$ denote the malware payload. Any extra instructions or payload exceeding $S_{\mathsf{SRAM}}$ cannot fit in fast memory and must be spilled into slower storage (off‐chip DRAM or secondary flash).  Formally, if the adversary’s memory footprint is
\[
S_{\mathsf{adv}} \;=\; S_{\mathsf{SRAM}} + S_{\mathsf{M}},
\]
then $S_{\mathsf{M}} > 0$ bytes must reside in a slower tier with bandwidth $B_{\mathrm{slow}} \ll B_{\mathsf{SRAM}}$.  Each pass through the checkpointed region will therefore incur an additional time penalty
\[
\Delta T \;\ge\; \frac{S_{\mathsf{M}}}{B_{\mathrm{slow}}}\,,
\]
which accumulates over the $P$ randomized passes.

\medskip

\noindent\textbf{Detectable Timing Penalty.}  
Each off-chip memory access incurs a significant latency penalty compared to on-chip SRAM. Over $P$ passes, even a small amount of swapped data $S_{\mathsf{M}}$ accumulates into a total delay:
\[
P \times \Delta T
\;=\;
P \times \frac{S_{\mathsf{M}}}{B_{\mathrm{slow}}}
\;>\;
\delta_{\mathrm{noise}},
\]
where $\delta_{\mathrm{noise}}$ is the timing noise floor and $B_{\mathrm{slow}}$ is the bandwidth of the slower memory. Thus, any violation of the space bound $S_{\mathsf{SRAM}}$ necessarily produces a detectable timing anomaly.

\medskip

\noindent\textbf{Implications for Attackers.}  
An adversary cannot both (a) execute extra malicious instructions or store additional data and (b) remain entirely within $S_{\mathsf{SRAM}}$.  Any attempt to remain stealthy will require off‐chip fetches. In every case, exceeding the on‐chip space budget forces storage swaps that our \textsc{MultiPass} timing measurement will expose.  This space–time enforcement transforms our challenge into a provably unconditional security check: attackers who violate the minimal space–time bounds of Horner’s evaluation are caught by latency penalties that cannot be masked.

\point{Dynamic Memory Access}  
By XOR’ing each accessed word \(v[i]\) with its corresponding coefficient \(s[i]\) before Horner‐rule accumulation, we prevent trivial memory regions from being ``skipped'' or optimized away.  Every word contributes a nontrivial term, forcing the DUT to execute the same sequence of loads and arithmetic operations on every pass.

\vspace{1ex}
Together, these design elements ensure that \TC{}’ provides an end‐to‐end, information‐theoretic guarantee: any unauthorized instruction or off‐chip fetch in the scanned region necessarily incurs a cumulative delay that the verifier will detect, enabling reliable rollback.

\subsection{Hardware Support Requirements}
\label{sec:hw-requirements}

While \TC{} shows that unconditional software root of trust can be established on commodity SoCs, achieving low‐variance, high‐confidence timing in practice ultimately demands minimal hardware support.  At a high level, three capabilities are essential:

First, the CPU clock domain must be held at a fixed frequency during the challenge.  Dynamic frequency scaling (DVFS) and thermal throttling introduce unpredictable slow‐downs that directly erode the tight timing margins on which our detector relies.  An external frequency‐lock mechanism, whether a dedicated PLL controller or a hardware jumper, ensures that every pass of the randomized‐polynomial loop proceeds at exactly the same rate.

Second, all other masters on the system interconnect must be quarantined or held in reset.  Background DMA, peripheral‐driven bus traffic, or timer interrupts can inject stray cycles and widen the noise floor.  By gating off non‐essential peripherals (e.g.\ disabling USB, NIC, MMC controllers) and freezing their requestors, the challenge traffic encounters a quiescent bus, so every observed delay truly reflects the DUT’s own memory and compute work. 

Third, the challenge–response link itself must exhibit sub‐microsecond, ideally nanosecond‐scale, jitter.  UART over an OS interrupt cannot deliver that; instead, a low‐latency channel toggled handshake line lets the verifier stamp start and end with minimal uncertainty.

In addition to these functional blocks, it matters deeply that the entire interface be open and auditable.  Any “black‐box” vendor IP on the challenge path becomes a single point of undetectable compromise: a stealth clock glitch generator or a hidden peripheral bridge could subvert the protocol. Therefore, minimal trusted platforms require fully open hardware specifications, as exemplified by Raptor's TALOS II, a PowerPC-based system designed specifically for high-assurance computing with complete firmware and hardware auditability \cite{raptor}. For the clock‐lock, bus‐quiesce, and low‐jitter timing unit.  While this shifts trust to the silicon vendor and fabricator, an open design allows independent audit and re‐use across products, mitigating supply‐chain risks.

\section{Prototype}
\label{sec:prototype}

We implemented \TC{} on the RockPro64 development board, which features a Rockchip RK3399 SoC with a quad-core ARM Cortex-A53 cluster. We chose an ARM core because it is representative of the processors used in a wide range of embedded and IoT devices today. Our \textsc{MultiPass} algorithm, which was hand-written in AArch64 assembly for implementation assurance, lives in the BL31 stage of ARM Trusted Firmware-A (TF-A). We expose it via two new Secure Monitor Call (SMC) interfaces, \texttt{checkpoint\_record} and \texttt{checkpoint\_replay}, so that any EL1/EL2 payload (including a guest OS under Hafnium) can invoke our low-level challenge logic in EL3.

\subsection{Hypervisor Integration}
We use Hafnium\cite{hafnium}, a lightweight Type-1 hypervisor for AArch64, to mediate record/replay between the guest and the secure monitor.  In Hafnium’s main hypercall handler we register two new function IDs, \texttt{HF\_RECORD\_CHECKPOINT} and \texttt{HF\_REPLAY\_CHECKPOINT}.  On receipt of \texttt{HF\_RECORD\_CHECKPOINT}, Hafnium:
\begin{enumerate}
  \item Flushes all on-chip state (TLBs, caches, pending workqueues, etc.).
  \item Packages the VCPU registers and the designated guest-memory range.
  \item Issues an SMC into BL31 invoking our \texttt{checkpoint\_record} routine.
\end{enumerate}
The \texttt{HF\_REPLAY\_CHECKPOINT} path performs the symmetric ``restore'' operation.  By placing these hooks in EL2, we achieve an in-band, record/replay mechanism that requires no additional firmware layers.

\subsection{Kernel Driver and Userland Interface}
On the host side we implement a Linux platform driver that registers a character device \texttt{/dev/tc}.  At probe time, the driver:
\begin{itemize}
  \item Locates a reserved DRAM region via Device-Tree and \texttt{ioremap()}s it.
  \item Pins execution to CPU 0 and hot-unplugs all other cores.
  \item Exposes two ioctls, \texttt{CHECKPOINT\_RECORD} and \texttt{CHECKPOINT\_REPLAY}, which wrap \texttt{stop\_machine()} contexts around the appropriate hypervisor calls and cache/TLB flushes.
\end{itemize}
From user space, recording a checkpoint is simply:
\begin{verbatim}        
    ioctl(fd, CHECKPOINT_RECORD, 0);         
\end{verbatim}
and replaying it is:
\begin{verbatim}
    ioctl(fd, CHECKPOINT_REPLAY, 0);
\end{verbatim}

\subsection{On-Demand Permutation Generation}
We must prevent attackers from offline-analyzing a fixed access order, because the checkpointed memory image is public. We therefore generate each permutation index on demand using a tiny Feistel-based block cipher keyed by the verifier’s seed.  Let $n$ be the number of elements and $b=\lceil\log_2 n\rceil$.  To produce the $i$th index, we:
\begin{enumerate}
  \item Zero-pad $i$ to a $b$-bit block.
  \item Run a small number of Feistel rounds (using a lightweight hash as the round function).
  \item If the result $j\ge n$, re-encrypt until $j<n$ (expected $<2$ tries).
\end{enumerate}
This construction uses $O(1)$ extra space, $O(1)$ expected time per index, and is invertible if needed.

\subsection{Secure Monitor Assembly Routine}
Our AArch64 assembly routine begins by saving all callee-saved registers, loading the prime modulus and Horner coefficients, and invoking \texttt{perm\_new} to seed the permutation generator.  The core loop is a nested two-level construct: an outer pass counter and an inner index loop.  Each iteration:
\begin{itemize}
  \item Calls \texttt{perm\_get()} to obtain the next randomized address.
  \item Performs a 64-bit load from the on-chip SRAM slice.
  \item Executes Horner’s rule using \texttt{mul}/\texttt{umulh} for 128-bit modular multiplication, \texttt{add}/\texttt{cmp}/\texttt{sub} for modular reduction, and \texttt{eor} to XOR with the coefficients computed from the random values (as in \ref{coff}).
\end{itemize}
All intermediate values remain in registers and the code is carefully aligned to prevent cache-line artifacts that could mask genuine DRAM misses.  After completing the prescribed number of passes, the routine restores registers and returns the accumulator via a secure output channel to the external verifier.

\subsection{Timing Channel}
We deploy a Raspberry Pi Pico W (RP2040) as our out-of-band timing monitor.  The Pico W listens on a UART link to the DUT and records timestamps with microsecond resolution.  By comparing the observed latency against the baseline, the verifier can detect any deviations caused by forced off-chip accesses or extra instructions.

\subsection{Challenge Generation}
Challenges are produced on a separate Rockpro64 acting as the ``challenge processor''. For each challenge session it samples fresh randomness for the $k+1$ Horner coefficients, the field element $x$, and the permutation seed $\pi$, and dispatches these parameters to the verifier MCU. Because each seed is unique and the permutation is generated on demand, the DUT cannot predict or prefetch future memory addresses.

After establishing a baseline timing distribution during calibration, the verifier detects tampering by:

\begin{enumerate}
\item Issuing a challenge and measuring the DUT's execution time
\item Applying appropriate statistical tests (e.g., z-score, modified z-score, etc.) to determine the probability that this timing came from the baseline distribution
\item If the confidence level is insufficient, the verifier has two options:
    \begin{itemize}
    \item Run additional independent challenges to reduce the probability of false positives
    \item Increase the number of passes $P$ to amplify timing differences, making anomalies more statistically significant
    \end{itemize}
\end{enumerate}

The choice of statistical test, significance threshold, and number of passes depends on the deployment's security requirements and acceptable false positive rate.

\subsection{Engineering Results}

\begin{figure}
  \centerline{\begin{tabular}{c}
  \colfig{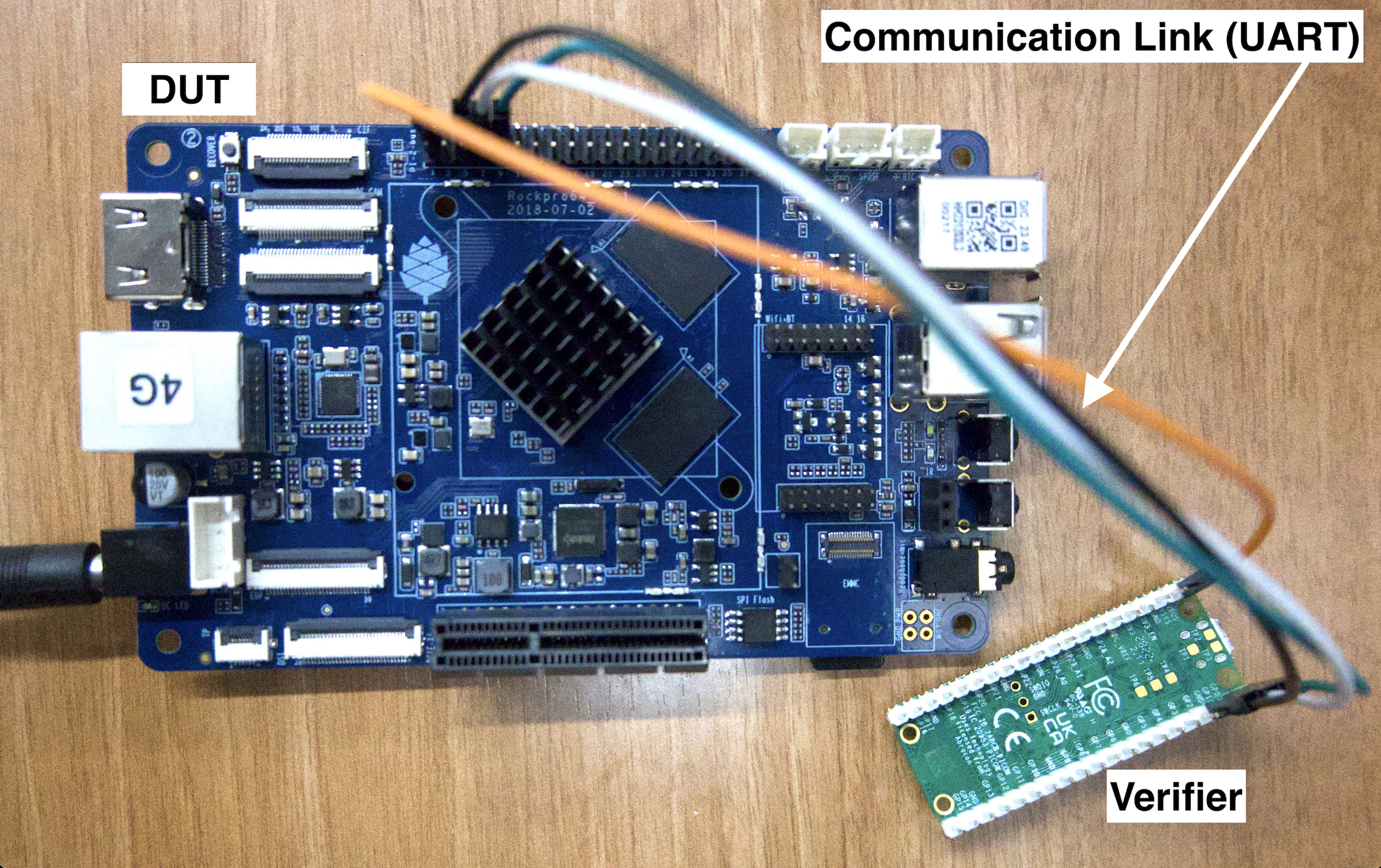} \\
  (a) Hardware \\
  \colfig{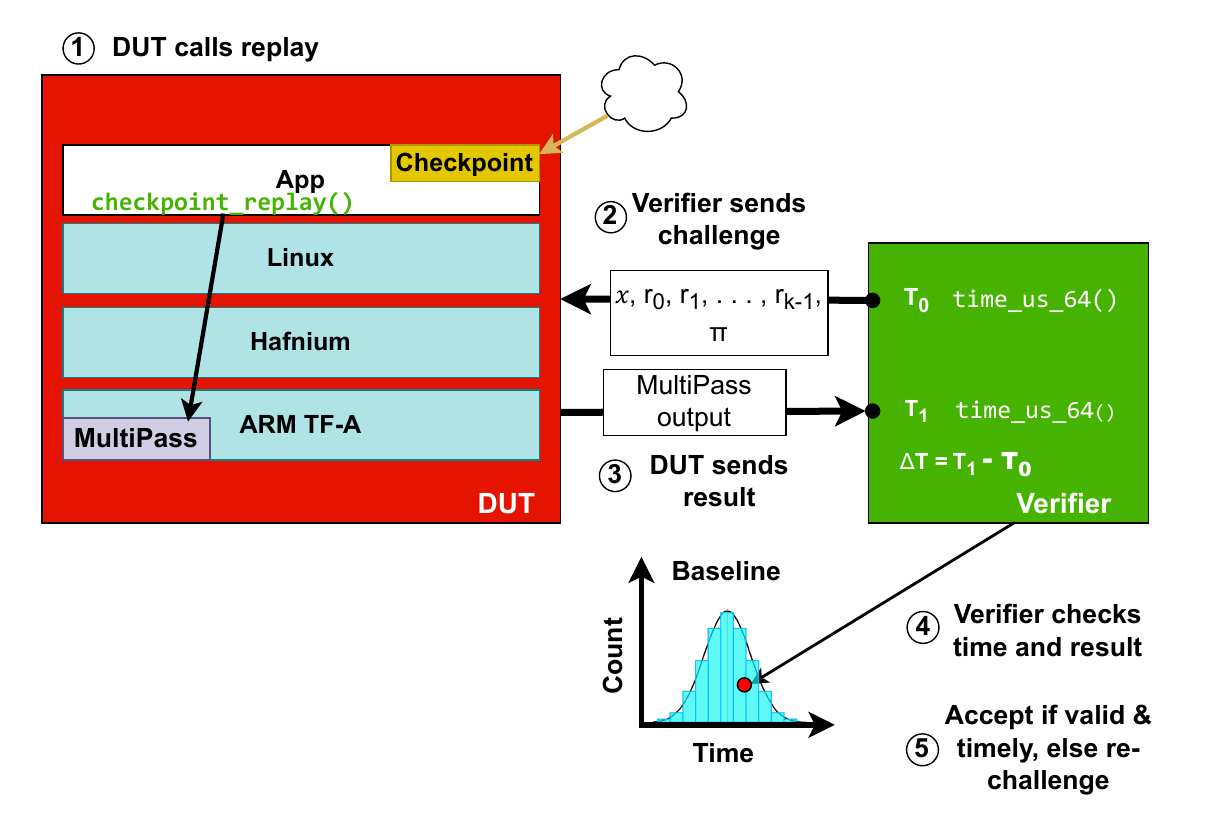} \\
  (b) Implementation Structure \\
  \end{tabular}}
  \caption{\TC{} prototype with RockPro64 (DUT) and Raspberry Pi Pico (Verifier)}
  
  \label{fig:arch}
\end{figure}

Figure~\ref{fig:arch} depicts our hardware/software prototype in the context of the architectural description above. This implementation requires no hardware modifications; we added 1,345 lines of C code to Hafnium, 274 lines of C code to Linux, and 1,464 lines of C code plus 127 lines of assembly to ARM TF-A.

\point{Portability Considerations}
The \textsc{MultiPass} algorithm is architecturally agnostic as it requires only basic arithmetic operations available on most processors. The timing-based detection remains effective on any platform where checked memory (SRAM/DRAM) exhibits lower latency than unchecked storage (flash/disk). A universal characteristic of modern memory hierarchies. While our current implementation uses hardware virtualization for checkpoint/restore, systems without such features would need alternative state management strategies. The core challenge logic remains portable, with only platform-specific checkpoint mechanisms requiring adaptation.

\section{Generalizable Practical Challenges}
\label{sec:practical-challenges}

In bringing \TC{} from theory to practice on commodity ARM hardware, we encountered and addressed a number of non-trivial engineering challenges.  These are important for understanding our prototype and we believe are generalizable to similar work.

\subsection{Moving Baseline and Nondeterminism}
\label{sec:moving-baseline}

In practice, the “clean” timing baseline for our 500‐pass scan is not perfectly static; several environmental and operational factors can shift it:

\begin{itemize}
  \item \textbf{Power‐level and DVFS changes.}  On many SBCs the CPU frequency (and core voltage) is adjusted dynamically to save power or respond to system load.  If the board’s governor selects a different P‐state than during calibration, the scan runtime will shift proportionally.  
  \item \textbf{Thermal throttling.}  Under sustained load the SoC temperature may exceed its safe operating point, triggering clock down‐throttling in hardware.  A scan performed immediately after, or during, such thermal events can be tens of milliseconds slower.  
  \item \textbf{PLL and clock jitter.}  Variations in the board’s crystal oscillator, phase‐locked loops, or supply voltage noise can introduce small cycle‐to‐cycle timing jitter that accumulates over long scans.  
  \item \textbf{Peripheral activity.}  Although we pin execution to CPU 0 and disable interrupts, background DMA, ECC scrubbers, or power‐management controllers may sporadically seize the memory bus, slightly perturbing the scan time.  
\end{itemize}

\noindent
\textbf{Mitigation.}  To maintain a stable baseline we recommend:
\begin{enumerate}
  \item \emph{Fixed P‐state:} lock the CPU to a known frequency and disable dynamic scaling during both calibration and the challenge.
  \item \emph{Thermal control:} attach a heatsink (or fan) and allow the processor cool before running the challenge. 
  \item \emph{Periodic recalibration:} rerun the baseline scan whenever the board’s power, cooling, or firmware configuration changes.
    \item \emph{Peripheral interference:} power off peripherals before running the challenge.
\end{enumerate}

\subsection{High-Precision Timing}

Reliable detection of off‐chip detours hinges on our ability to distinguish the tiny extra delays introduced by malicious memory accesses from the background jitter of the timing channel. In our prototype, we rely on the RP2040’s microsecond‐resolution timer over a UART link, whose intrinsic jitter, arising from interrupt latency and baud‐rate granularity, can be on the order of several microseconds. To overcome this coarse granularity, we iteratively increased the number of passes until a statistical test produced a significant separation between the baseline histogram and the attack histogram that used the next fastest available memory (see~\ref{experiment}). In an ideal instantiation, one would employ a low‐variability channel or a timer with picosecond resolution. With sub‐nanosecond resolution, the same randomized‐polynomial protocol could run with far fewer passes, reducing runtime overhead while still detecting any out‐of‐band memory or compute detours.

\subsection{Challenging Entire Memory State vs.\ Bootstrapping Trust}

Performing a full challenge over the entire 4\,GB of DRAM on a RockPro64 requires roughly 30 minutes on a single Cortex-A53 core. To reduce this latency, we employ a two-stage, “bootstrap” approach:

\begin{enumerate}
  \item \textbf{Test SRAM:}  
    First, restore and verify the contents of on-chip SRAM, which now holds the hash of the full checkpoint. Then, use this verified hash to authenticate the full checkpoint stored in DRAM.
  \item \textbf{DRAM bootstrapping:}  
    Once the contents of SRAM are verified and trusted, the code stored in SRAM can safely hash the checkpoint image in DRAM and zero out all unused memory. 
\end{enumerate}

\subsection{Non-maskable Interrupts (NMIs)}

Non‐maskable interrupts bypass all software level masking mechanisms. They always preempt the CPU. On many commodity SoCs, NMIs cannot be disabled or rerouted, so any NMI during the challenge window invalidates the timing measurement. Because NMIs are event‐driven and exceedingly rare on a healthy system, there is no fixed probability model for their occurrence. In practice, \TC{} simply retries the challenge.

We note that the prevelance of system management interrupts (SMIs) is likely to be a particular challenge on any modern x86 platform.   SMIs are generally established to run in firmware at boot time and then locked away.  Unfortunately, this firmware is largely a black box, making it both a source of potential malware, and impossible to validate even if not.

\subsection{Compression and Predictable Sequences}

To prevent an attacker from exploiting deterministic access patterns, e.g.\ by prefetching or caching predictable memory regions to avoid slow fetches, we randomize the order in which memory is traversed during the challenge.  Additionally, we compress the checkpoint data to minimize their footprint. This eliminates unused slack space that an adversary could use to stage malicious payloads and avoid the swaps to slower memory.

\subsection{CPU Microarchitecture Considerations}
\label{sec:cpu-microarch}

The RK3399’s Cortex-A53 cores feature an in-order, dual-issue pipeline with a small branch predictor, a minimal return stack, and limited nonblocking loads~\cite{arm_cortex_a53_trm}. None of which allow an attacker to hide extra work within our tight timing window.  In our design:

\begin{itemize}
  \item The Horner loop is a strict data-dependency chain: each 64-bit modular multiply-add depends on the previous result.  Since the A53 cannot reorder or split that sequence without corrupting the polynomial, any interleaved instructions stall on the single address generator or the ALU complex port.
  \item We flush instruction and data caches before each challenge, so every SRAM access and any forced DRAM/off-chip detour pays full latency.  With only eight nonblocking-load entries, injected memory operations quickly saturate the buffer and lengthen execution time.
  \item The single load/store unit and in-order writeback stage force any extra load/store cycles or execution-latency interlocks (e.g., waiting for a 4-cycle multiply) to serialize, directly adding to the measured runtime.
  \item Interrupts and the MMU are disabled during measurement, preventing asynchronous events or prefetchers from skewing timing.
\end{itemize}

Together, these properties ensure that no microarchitectural trick on a Cortex-A53 can hide the cost of extra instructions or off-chip fetches: either the polynomial result is invalid or the runtime exceeds the calibrated bound, allowing the verifier to detect tampering.

\point{Advanced Processors}
Out-of-order CPUs with larger reorder buffers, multiple execution units, and aggressive speculation cannot break \TC{}'s security guarantees due to the fundamental structure of our algorithm. \textsc{MultiPass} computes a serial accumulator where each iteration depends on the previous result and fetches from a pseudorandom address determined by that result. While OoO execution can hide small latencies (e.g., precomputing the next address in the permutation), it cannot break this dependency chain or issue multiple permuted memory fetches in parallel. The throughput remains fundamentally bounded at one word per iteration, regardless of the CPU's out-of-order capabilities. The serial data dependency in our polynomial evaluation ensures that even the most aggressive speculation cannot create exploitable idle cycles for malicious code to execute undetected.

\subsection{DMA TOCTOU Vulnerabilities}
\label{sec:dma}
Direct Memory Access (DMA) enables peripherals like network cards, storage controllers, and GPUs to transfer data without CPU involvement. In principle, an attacker controlling a malicious peripheral (or compromising legitimate peripheral firmware) could attempt to queue DMA transfers to overwrite memory regions after verification. \TC{} addresses this threat through multiple enforcement strategies that bound DMA effects during the challenge window:

\textbf{(1) DMA Descriptor Challenge:} We stop DMA queues and zero DMA descriptors, which reside in DRAM. By including these descriptor locations in the challenged region, any malicious DMA requests become immediately visible and cause challenge failure.

\textbf{(2) Peripheral Quiescence:} Deployers can hard-quiesce peripherals by asserting reset signals, gating clocks, disabling DMA engines/channels, and masking interrupts. The relevant control/status registers are included in the challanged region to verify and maintain this quiesced state throughout the challenge.

\textbf{(3) SRAM Isolation:} In SRAM-bootstrap mode, \TC{} leverages the fact that on many SoCs, on-chip SRAM is not DMA-accessible, providing natural isolation from DMA-based attacks.

\TC{}'s security guarantee holds when at least one of these enforcement strategies is active during the challenge window.

\subsection{Do We Have the Fastest Implementation?}
Our \textsc{MultiPass} implementation achieves the theoretical minimum for sequential polynomial evaluation: exactly $d$ fetches, $d$ multiplies, and $d$ adds per pass. On the A53, this translates to one multiply, one modular reduction sequence, one load, and one XOR per iteration. Our hand-tuned assembly saturates both issue slots with zero wasted cycles—the strict dependency chain prevents any reordering or parallelization.

Could vectorization help? Vectorization cannot improve our performance due to the fundamental structure of our algorithm, not architecture-specific limitations. Our loop computes a serial accumulator where each iteration depends on the previous result and fetches from a pseudorandom address determined by that result. SIMD lanes have no independent work to parallelize, each step must complete before the next can begin. Attempting vectorization would only add overhead from packing/unpacking operations and GPR-to-SIMD register moves without any throughput benefit. 

The engineering challenges we encountered—firmware integration, noise calibration, memory shuffling, and microcontroller coordination are solvable implementation details, not fundamental barriers. Our working prototype demonstrates the potential for establishing an unconditional software root of trust on commodity hardware, paving the way for future exploration and development.

\section{Evaluation}
Is it possible to programmatically detect malware with high confidence using the proposed technique? After $n$ passes what is the probability that we would not be able to detect a single malware instruction? Our detection system must have a very low false negative rate and low false positive rate.

\subsection{Detecting a Single-Instruction}

The fundamental challenge in \TC{} is detecting minimal adversarial interference, specifically, the injection of even a single malicious instruction. Unlike traditional malware that requires substantial code footprints, a sophisticated adversary might attempt to hide just one critical instruction that redirects control flow or leaks sensitive data. This represents the hardest detection scenario: the adversary minimizes their footprint while maximizing their impact.

To retain even this single instruction in memory scanned by \TC{}, an adversary must evict part of the checked state to make room. Since \TC{} fully occupies the challenged region during verification, any such evasion requires swapping data between fast, challenged memory and slower, unmonitored memory. The minimum viable attack requires at least two instructions per word:

\begin{verbatim}
LDR Xn, [source_address]
STR Xn, [dest_address]
\end{verbatim}

Each swap operation incurs a timing penalty that accumulates across the challenge. Our evaluation demonstrates that \TC{} can detect these minimal timing perturbations with high confidence.

\subsection{Experimental Setup}
\label{experiment}

We evaluate \TC{}'s detection capability on a RockPro64 board with an RK3399 SoC, using a single Cortex-A53 core to test 192~KB of on-chip SRAM (INTMEM0). We selected 500 passes based on preliminary testing that showed this number provides sufficient statistical separation between baseline and attack distributions while maintaining reasonable challenge time (9.6 seconds). With more passes, the increase in challenge time offers diminishing detection benefits as the distributions are already well-separated. Our experiments measure end-to-end latency across 500 randomized polynomial scans under three scenarios:

\begin{enumerate}
  \item \textbf{Baseline:} Standard \textsc{MultiPass} execution with no adversarial activity
  \item \textbf{DRAM Attack:} Adversary swaps one word to off-chip DRAM per pass
  \item \textbf{IOMEM Attack:} Adversary swaps one word to memory-mapped I/O space per pass
\end{enumerate}

Each scenario was executed 50 times to characterize timing distributions. We disabled interrupts to ensure timing consistency while maintaining realistic system conditions with background peripherals active.

\subsection{Data Independence}

\begin{figure}[ht]
  \centering
  \colfig{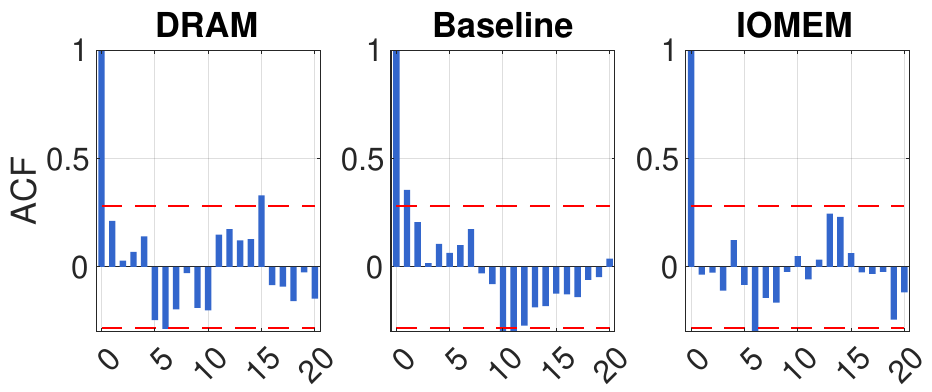}
  \caption{Our data shows no serial correlation, suggesting independence-assuming statistical tests apply. The dotted red lines represent the 95\% confidence interval for the noise floor.}
  \label{fig:sercor}
\end{figure}

In order for statistical comparisons to be made a common requirements is that our sample is independent. Figure~\ref{fig:sercor} shows the serial correlation plots for all three of our scenarios along with bounds on signification correlations. As can be seen the data has no serial correlation with high confidence. 

\subsection{Empirical Comparisons}

\begin{figure}
  \centering
  \colfig{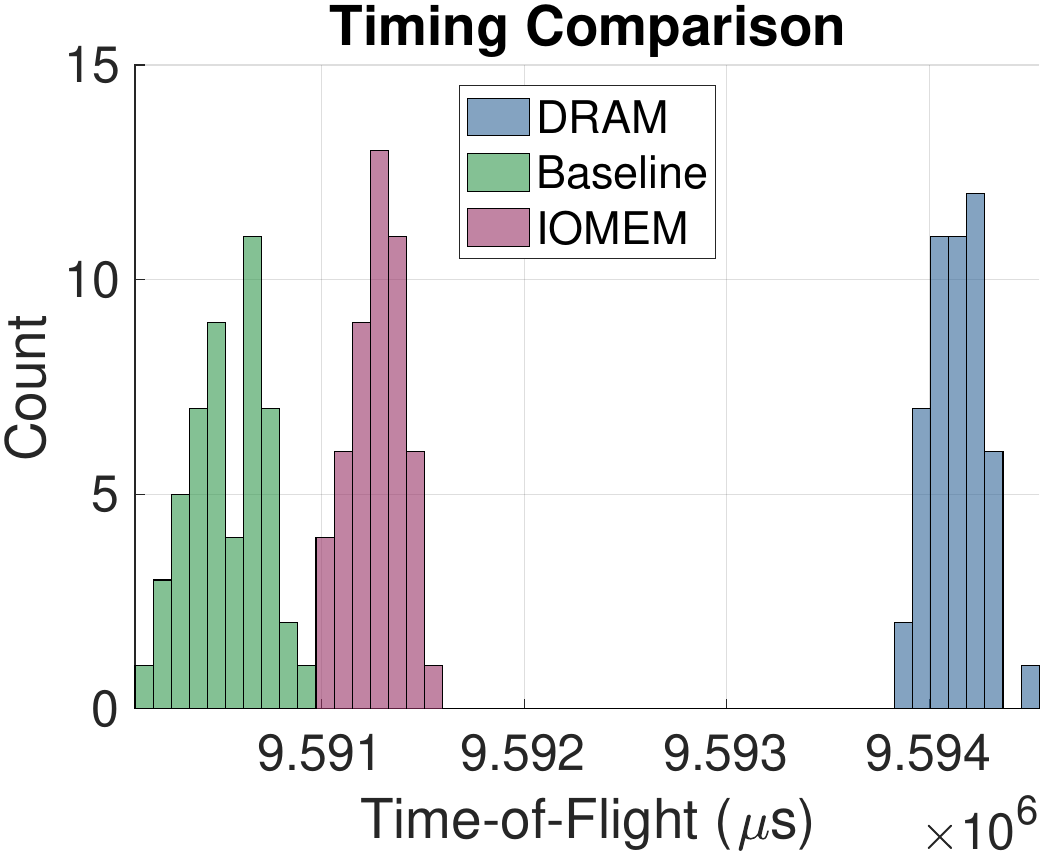}
  \caption{Empirical distributions show a clear separation between baseline and attack scenarios. Even the fastest IOMEM attack is readily distinguishable from legitimate execution.}
  \label{fig:combined}
\end{figure}

Figure~\ref{fig:combined} shows the histograms of the baseline, DRAM attack, and IOMEM attack scenarios.  Given the lack of serial correlation, this is sufficient representation of the data.

\point{Baseline}
The baseline distribution is for legitimate challenge runs. The measurements exhibit a tight distribution with mean 9.590$\times 10^6$~$\mu$s and standard deviation 185~$\mu$s. This establishes our detection threshold: any execution exceeding this baseline by multiple standard deviations indicates potential tampering.

\point{DRAM-Based Attacks}
The most detectable attack vector involves swapping data to off-chip DRAM. As shown in Figure~\ref{fig:combined}, DRAM-based evasion produces a clear rightward shift in the timing distribution. The mean execution time increases to 9.594$\times 10^6$~$\mu$s—a small but consistent delay of approximately 4,000~$\mu$s compared to the baseline.

\point{IOMEM-Based Attacks}
A more sophisticated adversary might attempt to use memory-mapped I/O regions, which typically offer lower latency than DRAM. Figure~\ref{fig:combined} shows that IOMEM attacks produce a smaller but still detectable timing shift. The mean execution time of 9.591$\times 10^6$~$\mu$s represents only a 1,000~$\mu$s delay—one quarter of the DRAM overhead compared to the baseline.

\subsection{Statistical Comparisons}

We now examine the statistical divergence between baseline and attack distributions to confirm that single-instruction attacks create distinct measurements that are fundamentally different from legitimate execution.

We use two different tests to do so:
\begin{itemize}
\item The unpaired t-test~\cite{T-TEST} compares the means of the distributions with the null hypothesis that they are the same.  
\item The Kolmogorov-Smirnov test~\cite{KS-TEST} compares two distributions with the null hypothesis that they are the same.
\end{itemize}

\begin{figure*}
  \centering
  \begin{tabular}{lrrrrr}
    \toprule
    \textbf{Scenario} & \textbf{Mean} & 
    \textbf{Std Dev} & \textbf{t-test} & \textbf{ks-test}  \\
    & ($\times 10^6~\mu$s) & ($\times 10^3~\mu$s) & \textbf{p-value} & \textbf{p-value} \\  
    \midrule
    Baseline     & 9.591 & 0.185 & -- & -- \\ 
    DRAM Attack  & 9.594 & 0.130 & $1.67 \times 10^{-105}$ & $2.16 \times 10^{-23}$ \\
    IOMEM Attack & 9.591 & 0.128 & $1.86 \times 10^{-41}$  & $2.16 \times 10^{-23}$ \\ 
    \bottomrule
  \end{tabular}
  \caption{Both attack types are distinguishable from the baseline. All statistical tests indicate divergence, with extremely low p-values from the t-test and KS-test. Chebyshev bounds with $k = 31.6$ also flag the attacks as outliers relative to the baseline.}
  
  \label{fig:stats}
\end{figure*}

\begin{figure*}[t]
  \centering
  \begin{tabular}{lrrrrr}
    \toprule
    \textbf{Scenario} & \textbf{Mean} & 
    \textbf{Std Dev} & \textbf{t-test} & \textbf{ks-test} \\ 
    & ($\times 10^6~\mu$s) & ($\times 10^3~\mu$s) & \textbf{p-value} & \textbf{p-value} \\ 
    \midrule
    Baseline (Full)     & 1731.895 & 16.543 & -- & -- & \\ 
    MMC Attack          & 1735.465 & 28.587 & $4.37 \times 10^{-4}$ & $9.70 \times 10^{-2}$ \\ 
    \bottomrule
  \end{tabular}
  \caption{Full-memory challenge results demonstrate that MMC attack timing remains statistically distinguishable from baseline, validating that our detection methodology scales to gigabyte-sized memory regions.}
  \label{fig:stats-full}
\end{figure*}

Figure~\ref{fig:stats} presents three complementary statistical tests that show that the distributions are significantly different. The extremely low p-values ($< 10^{-23}$) indicate that the probability of observing such timing differences by chance alone is small under the null hypothesis that attacks and baseline come from the same distribution.

\subsection{Detecting Attacks}
While demonstrating that attack distributions differ from baseline is important, deployment requires the verifier to classify individual timing measurements as either legitimate or malicious. We evaluated three statistical approaches for single-point outlier detection by testing each individual data point from both the baseline and attack distributions, simulating how a verifier would classify measurements.

We used the following detection methods:
\\

\textbf{Percentile-Based Detection (Non-parametric)~\cite{Z}:} This method makes no distributional assumptions and directly uses the empirical baseline distribution. Each test point is compared against the baseline data to determine its percentile rank. This approach is robust to non-normal distributions.

\textbf{Z-Score Detection (Parametric)~\cite{Z}:} The classical z-score method assumes the baseline follows a normal distribution. For each test point, we compute $z = (x - \mu)/\sigma$, where $\mu$ and $\sigma$ are the baseline mean and standard deviation.

\textbf{Modified Z-Score Detection (Non-parametric)~\cite{MAD-Z}:} This method replaces the mean and standard deviation with robust statistics, the median and median absolute deviation (MAD), making it less sensitive to outliers and applicable to non-normal distributions.

The detection thresholds for each method represent standard cutoffs in statistical outlier detection. For the percentile-based method, the 2.5\% threshold means we flag points falling below the 2.5th or above the 97.5th percentile of the baseline distribution, a common choice that captures 95\% of normal behavior while identifying extremes in both tails. The z-score method's 3$\sigma$ threshold is the classical ``three-sigma rule,'' which assumes that 99.7\% of normally distributed data falls within three standard deviations of the mean; points beyond this are considered statistical anomalies. The modified z-score threshold of 2.5 is more liberal than the commonly used 3.5, but was chosen based on empirical optimization, it provided the best balance between detection sensitivity and false positive rates in our experiments.

\point{Detection Performance}
For each detection method we used the following thresholds: 

\begin{itemize}
   
    \item \textbf{Percentile-based:} We identified extreme values falling below the 2.5th percentile or above the 97.5th percentile of the baseline distribution.
      \item \textbf{Regular z-score:} We used a threshold of $\lvert z \rvert > 2$.
       \item \textbf{Modified z-score:} We flagged outliers where $\lvert z \rvert > 2.5$.
  
\end{itemize}

 For the percentile-based method, the 2.5\% and 97.5\% cutoffs capture the most extreme 5\% of data points, aligning with a typical 95\% confidence interval for normal behavior. The regular z-score threshold of 2 corresponds to approximately 95\% coverage under a Gaussian distribution.

The modified z-score method uses the median and median absolute deviation (MAD) for robustness against skewed or non-Gaussian distributions. We empirically determined that $\lvert z \rvert > 2.5$ offered the best trade-off between sensitivity and false positive rate in our evaluations.

The results in Figure~\ref{fig:detection-performance} confirm that \TC{} can reliably distinguish even minimal single-instruction attacks from legitimate execution with high probability. We achieve the low false negative rate and the false positive rates that are necessary to make this method deployable. This is what was achieved with one challenge, but to achieve arbitrarily high confidence, the verifier can repeat challenge until the cumulative false negative probability falls below a target threshold.

\begin{figure}[h]
\centering
\begin{tabular}{lccc}
\toprule
\textbf{Method} & \textbf{FPR} & \textbf{FNR} \\
                & (\%) & (\%)  \\
\midrule
\multicolumn{4}{l}{\textit{DRAM Attack Detection}} \\
Percentile             & 0.0 & 0.0  \\
Z-Score                & 0.0 & 0.0  \\
Modified Z             & 0.0 & 0.0   \\
\midrule
\multicolumn{4}{l}{\textit{IOMEM Attack Detection}} \\
Percentile             & 0.0 & 0.0   \\
Z-Score                & 0.0 & 0.0   \\
Modified Z             & 9.0 & 0.0   \\
\bottomrule
\end{tabular}
\caption{Detection performance metrics. FPR: False Positive Rate, FNR: False Negative Rate.}
\label{fig:detection-performance}
\end{figure}

\subsection{Full-DRAM vs SRAM-Bootstrap Challenge}

We evaluated \TC{} across the entire system memory (192~KB SRAM + 4~GB DRAM) on a single core. A complete single-pass challenge requires approximately 28 minutes (1.73$\times 10^9$~$\mu$s).

To validate that our statistical detection methodology scales beyond SRAM-only challenge, we tested detection capability by introducing a minimal I/O operation (512-byte MMC read/write) when the scan index reaches 100. This attack produced timing deviations exceeding 200 standard deviations from baseline. The statistical tests confirm significant separation (t-test p-value $< 10^{-3}$), demonstrating that the same outlier detection techniques proven effective for SRAM-bootstrap challenge, whether percentile-based, z-score, or modified z-score, remain equally applicable at full-memory scale.

The fundamental principle remains unchanged: any off-chip memory access introduces measurable timing delays that accumulate across passes and exceed baseline variance. While we present only t-test and KS-test results here for brevity, the three detection methods analyzed in our SRAM experiments would yield similar detection performance, as the underlying timing perturbations follow the same physical constraints.

Though the challenge time makes frequent full-system checks impractical. In deployment, we recommend focusing on critical memory regions (e.g., kernel code, security monitors) for regular most challenge, with periodic full-system scans for comprehensive verification.

\section{Conclusion and Future Work}

We introduced \TC{}, the first unconditional trust framework for embedded and IoT devices that requires no stored secrets or trusted hardware and makes no assumptions about attacker capabilities. By recording CPU registers, SRAM, and DRAM checkpoints and applying $k$-independent randomized-polynomial evaluation via Horner’s rule, \TC{} forces any malicious code to incur detectable off-chip storage latency. Multiple randomized passes, measured by an external tamper-resistant microcontroller, ensure that even a single off-chip fetch exceeds the calibrated noise floor.

Our prototype on commodity ARM hardware (ARM Trusted Firmware-A and Hafnium) includes an SRAM-only evaluator and a full-memory mode. The SRAM-only configuration (500 passes over 192 KB) completes in roughly 9 s and reliably detects any extra instruction or off-chip access. These results show that keyless, information-theoretic root of trust establishment is possible on constrained platforms and provably secure against strong adversaries.

Future work includes exploring bandwidth-hard functions that tie challenge runtime directly to memory bandwidth, regardless of CPU parallelism, to further strengthen security. We also plan to investigate Processing-in-Memory architectures (e.g., UPMEM DPUs) to speed up challenges on larger memory sizes. We hope \TC{} inspires further research into unconditional security for next-generation connected devices.

\bibliographystyle{IEEEtran}
\bibliography{references}

\end{document}